\documentclass[aps,prb,showpacs,twocolumn,groupedaddress]{revtex4-1}
\usepackage{graphicx,graphics,color,epsfig}
\usepackage{amsmath}
\usepackage{amssymb}
\usepackage{amsfonts}
\usepackage{amsfonts}
\usepackage{epstopdf}
\usepackage{bm}
\usepackage{times,xspace}
\usepackage{color}

\begin{document}

\title{Adiabatic transformation as a search tool for new topological insulators: distorted ternary Li$_2$AgSb-class semiconductors and related compounds}

\author{Hsin Lin$^{1}$, Tanmoy Das$^{2}$, Yung Jui Wang$^{1}$, L.A. Wray$^{3,4}$, S.-Y. Xu$^{3}$, M. Z. Hasan$^{3}$, A. Bansil$^{1}$}
\affiliation{$^1$Department of Physics, Northeastern University, Boston, Massachusetts 02115, USA.\\
$^2$Theoretical Division, Los Alamos National Laboratory, Los Alamos, New Mexico 87545, USA.\\
$^3$Department of Physics, Joseph Henry Laboratories, Princeton University, Princeton, New Jersey 08544, USA.\\
$^4$Advanced Light Source, Lawrence Berkeley National Laboratory, Berkeley, California 94305, USA.
}


\date{\today}

\begin{abstract}
We demonstrate that first-principles based adiabatic continuation approach is a very powerful and efficient tool for constructing topological phase diagrams and locating non-trivial topological insulator materials.
Using this technique, we predict that the ternary intermetallic series Li$_2M'X$ where $M'$=Cu, Ag, Au, or Cd, and $X$=Sb, Bi, or Sn, hosts a number of topological insulators with remarkable functional variants and tunability. We also predict that several III-V semimetallic compounds are topologically non-trivial. We construct a topological phase diagram in the parameter space of the atomic numbers of atoms in Li$_2M'X$ compounds, which places a large number of topological materials presented in this work as well as in earlier studies within a single unified topological framework. Our results demonstrate the efficacy of adiabatic continuation as a useful tool for exploring topologically nontrivial alloying systems and for identifying new topological insulators even when the underlying lattice does not possess inversion symmetry, and the approaches based on parity analysis are not viable.
\end{abstract}


\maketitle

The recent discovery of topological insulators has opened up a new research direction in condensed-matter and materials science communities with tremendous potential for practical applications as well as for conceptual novelties.\cite{reviewHasan,reviewZhang,reviewMoore,wilchek} The unusual properties of topological insulators become manifest at their surfaces through the appearance of spin-polarized metallic states even though the bulk remains an insulator. These surface states have been predicted to host a large number of interesting topological phenomena.\cite{ZhangDyon,essinaxion,exciton,Zhangmonopole,wilchek,Majorana,WrayCuBiSe} For instance, the electrodynamics of a topological insulator entails additional terms in Maxwell's equations yielding novel Faraday-Kerr effect, magnetoelectric effect, and axion physics.\cite{ZhangDyon,essinaxion} Other exciting possibilities with topological surface states involve exciton condensation\cite{exciton}, presence of an image magnetic monopole induced by an electric charge, and Majorana fermions induced by the proximity effect from a superconductor.\cite{Zhangmonopole,Majorana,WrayCuBiSe} In this way, topological insulators may be thought of as unifying condensed matter physics with particle physics and cosmology.\cite{wilchek}

The practical realization of many of the aforementioned opportunities with topological insulators has not been possible to date because the choice of available topological materials is quite limited. The currently known topological insulators mainly belong to the $Z_2$ class, which includes Bi$_2$Te$_3$ family \cite{MatthewNatPhys,BiTeSbTe,tetradymite}, the ternary half-Heusler compounds \cite{heuslerhasan,heuslerZhang}, thallium-based chalcogenides \cite{TlBiTe2,TlBiTe2Zhang,SuYangThallium}, Ge$_m$Bi$_{2n}$Te$_{m+3n}$-class \cite{SuYangGBT}, quaternary chalcogenide compounds, and ternary famatinite compounds \cite{Ray2011}. The search has also been extended to other classes of materials where many-body physics and/or crystalline symmetries conspire to produce non-trivial topological phases.\cite{Tkondo,TAFM,TMott,THO,topocrystal} Despite this progress however the realization of wide classes of emergent topological properties has remained a challenge.

In connection with the materials discovery effort, adiabatic continuation is an elegant and powerful tool for obtaining a handle on the Z$_2$ time reversal invariant even when the lattice does not possess a center of inversion symmetry. The idea is to start with a known topological insulator and exploit the fact that if the Hamiltonian of this system can be transformed into that of the new system of interest through a continuous series of transformations without inducing a band inversion, then the new material will also be a topological insulator. In what follows, the adiabatic continuation approach will be seen to be not only a method for identifying the topological property of a material, but to also provide a pathway for finding new classes of topologically interesting materials. We specifically discuss how gray Sn, which is known to be topologically non-trivial, can be mutated to obtain new families of topologically non-trivial phases in zinc-blende-like materials and their superlattice variations. In particular, we predict that distorted-Li$_2$AgSb, a lightweight compound, is adiabatically connected to Sn, and thus harbors a 3D topological insulator state and that its cubic ground state lies near a critical point. In contrast, the related Li$_2$CuSb-type compounds are found to be topologically trivial. A topological phase diagram in the parameter space of the atomic numbers of the constituent elements in Li$_2$AgSb structure is also adduced, which places a number of topological materials within a uniform topological framework. The ternary Li$_2M'X$ series (with a number of variants) is thus identified as a new platform for deriving a host of topological compounds, alloys and nanoscale heterostructures via the flexibility of their lattice parameters and spin-orbit interaction strength. 
We also predict several III-V semimetallic compounds to be topologically nontrivial.

\begin{figure}
\includegraphics[width=8.5cm]{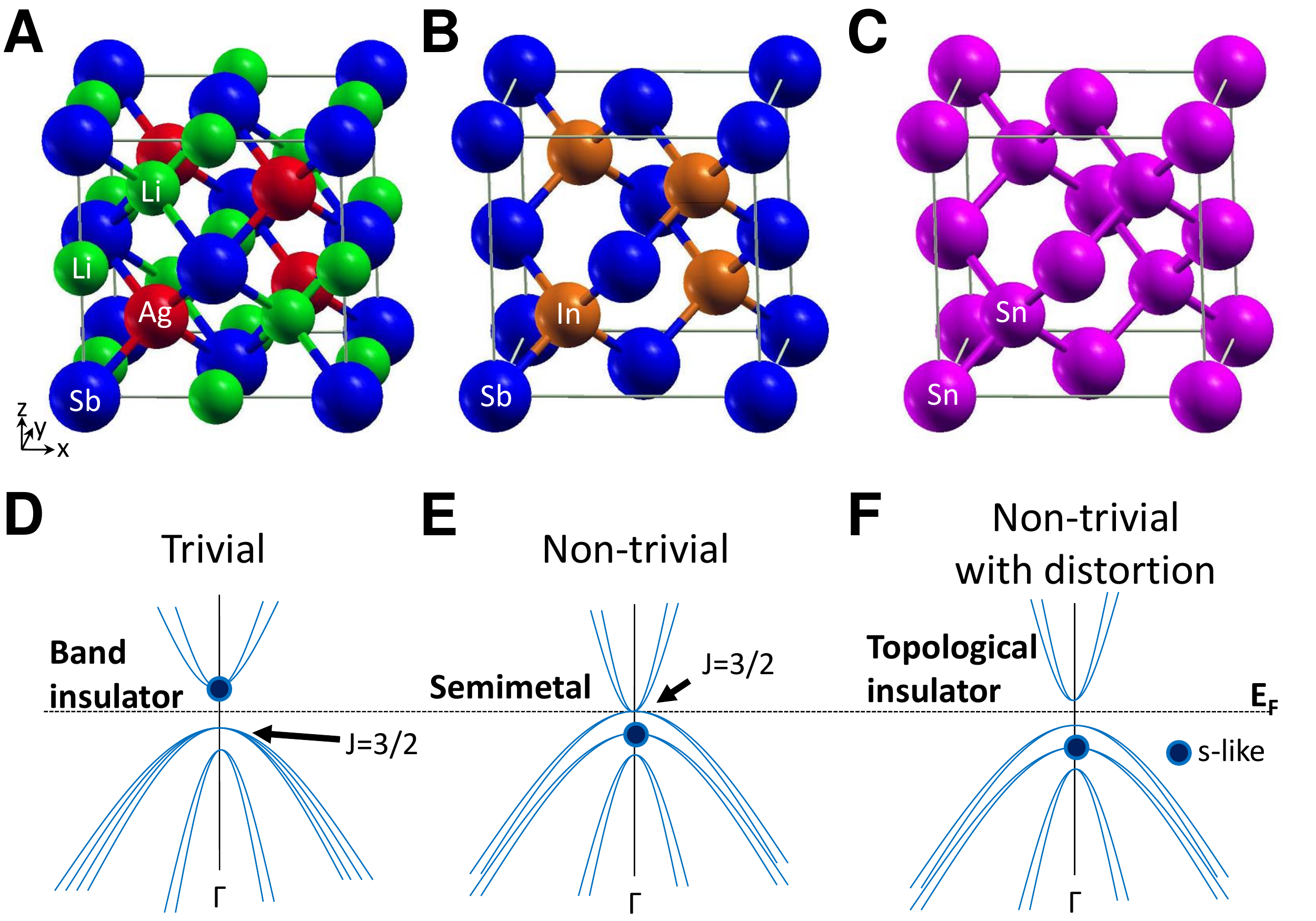}
\caption{\label{fig:sketch}
(Color online) {Crystal structure and topological band inversion.} (a) Crystal structure of Li$_2$AgSb. Li, Ag, and Sb are denoted by green, red, and blue balls, respectively. Sb and Ag form the zinc-blende sublattice. (b) The zinc-blende structure of InSb. In and Sb are denoted by gold and blue balls, respectively. (c) The diamond structure of gray tin. Diagrams in (d) (e), and (f) illustrate band structures near the $\Gamma$-point for trivial insulator, non-trivial semimetal, and non-trivial insulator, respectively. Blue dots denote the s-like orbitals at $\Gamma$. Band inversion occurs in the non-trivial case where the s-like orbitals at the $\Gamma$-point fall below the four-fold degenerate $j=3/2$ states. The degeneracy of the $j=3/2$ states is lifted by the lattice distortion in the non-cubic case.}
\end{figure}

We start our discussion of band inversion in the electronic structure with reference to gray tin\cite{FuKane}. It is useful to understand first how a non-trivial topological phase emerges in Sn, whereas its isostructural counterparts such as silicon (Si) and germanium (Ge) are topologically trivial, see Fig.~1(c). As shown in Fig.~1(d), the $s$-like conduction bands (blue dots) lie well above the $p$-like valence bands in Si and Ge, indicating the absence of band inversion in Si and Ge. As the lattice constant increases upon replacing Si or Ge with Sn, the conduction band moves down to touch the valence band, resulting in a band inversion at only the $\Gamma$-point. This pushes the twofold degenerate $s-$like bands, which were originally lying in the conduction band in Si or Ge, below the valence bands of the fourfold degenerate $p-$states of total angular momentum $j=3/2$, and the Fermi level ($E_F$) goes through the $p-$states as shown in Fig.~1(e). As the $s-$states are now occupied at $\Gamma-$point, the $Z_2$ topological invariant picks up an extra time-reversal factor of -1 compared to Si or Ge. Therefore, Sn is a topologically nontrivial zero gap semiconductor or semi-metal in its pristine phase.\cite{FuKane}

If we define band-inversion strength (BIS) as the energy difference between the $j=3/2$ states and the twofold degenerate $s$-states at the $\Gamma$-point, compounds with positive values of BIS will be topologically non-trivial, while those with negative values will be topologically trivial. The discussion in the preceding paragraph suggests a general route for realizing a topologically insulating phase where one lifts the 4-fold degeneracy of the $j=3/2$ states at the $\Gamma-$point by breaking the cubic symmetry of the crystal without inducing a negative band-inversion strength. We explore this possibility by showing how interplay of effects of finite distortion, heavy element substitution, and/or formation of superlattice of cubic structure, can either remove the band inversion if the band-inversion strength is not sufficiently large, or conversely, it can enhance the band-inversion strength.

In carrying out adiabatic transformations, it is important to consider the crystal structure first. Here, we begin with the ternary compositions Li$_2M'X$ ($M'$=Cu, Ag, and Au, $X$=Sb and Bi) whose crystal structure belongs to the space group $F\bar{4}3m$, with the atomic arrangement presented in Fig.~1(a). $M'$ and $X$ atoms occupy the Wyckoff 4$d$ and 4$a$ positions, respectively. Li atoms fill the remaining empty space in the Wyckoff 4$b$ and 4$c$ positions. Because $M'$ and $X$ atoms form a zincblende type sublattice, these materials resemble InSb if Li atoms are removed [Fig.~1(b)]. Note that In and Sb precede and follow Sn in the periodic table and InSb resembles gray Sn with diamond structure as shown in Fig.~1(c). These observations suggest that Li$_2M'X$ compounds could be candidates for Z$_2$ topological insulators if their electronic band structures resemble the band structure of structurally similar Sn. As already noted above, since there is no spatial inversion symmetry in Li$_2M'X$ or in zinc-blend InSb, parity methods\cite{FuKane} of computing $Z_2$ topological number are not viable.

\begin{figure}
\includegraphics[width=8.5cm]{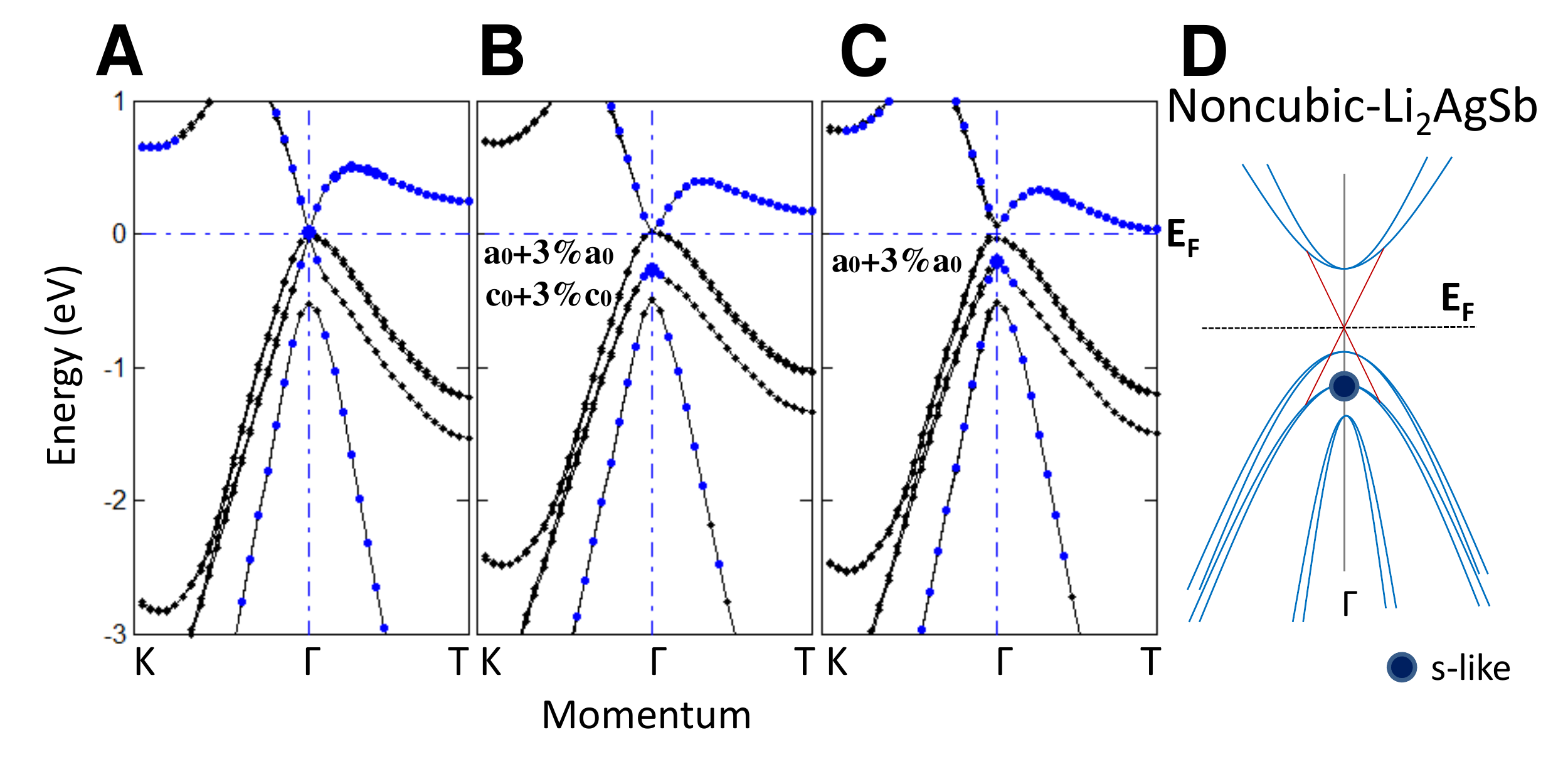}
\caption{ (Color online) {Electronic band structures of topological insulating state in Li$_2$AgSb with lattice distortion.}
(a) Li$_2$AgSb with experimental lattice constants. (b) Li$_2$AgSb with 3\% expansion of lattice constants. (c) Rhombohedral Li$_2$AgSb with hexagonal lattice constants $a=a_0+3\%a_0$ and $c=c_0$, where $a_0$ and $c_0$ correspond to the experimental values. (d) A sketch of band structure near the $\Gamma$-point for topologically non-trivial Li$_2$AgSb with a lattice distortion. The s-like states are marked with a blue dot. Lattice distortion causes a gap to open at $E_F$, resulting in a topological insulating state in which surface bands will span the bulk band gap, resembling the dispersion plotted with red lines.}
\end{figure}

First-principles band structure calculations were performed with the linear augmented-plane-wave (LAPW) method using the WIEN2K package \cite{wien2k} within the framework of the density functional theory (DFT). The generalized gradient approximation (GGA)\cite{PBE96} was used to describe the exchange-correlation potential. Spin orbital coupling (SOC) was included as a second variational step using a basis of scalar-relativistic eigenfunctions.

The computed band structure of Li$_2$AgSb along high symmetry lines in the Brillouin zone is presented in Fig.~2(a). Away from the $\Gamma-$ point, the Fermi level is completely gapped and thus the topological properties can be determined from observations of band structure only near the $\Gamma$-point. Focusing on the band structure very close to the Fermi level, we find that the orbital angular momentum symmetries of these compounds are
identical to those of the low energy spectrum in Figs.~1(d) and 1(e). The $s-$type and $p-$type states are nearly degenerate at the $\Gamma$-point around which the conduction bands and one of the valence band have almost linear dispersion, indicating that the system is at the critical point of the topological phase [Fig.~2(a)]. With a small expansion of lattice, the $s/p$ band inversion occurs as shown in Fig.~2(b). Conversely, the topological band inversion in Li$_2$AgSb can be removed altogether by uniformly decreasing all lattice constants, demonstrating chemical tunability of the system.

Since Li$_2$AgSb is at the critical point of the topological phase, it is possible to achieve a 3D topological phase by a finite distortion which lifts the degeneracy of the $j=3/2$ states and simultaneously induces a band inversion. We find that a rhombohedral distortion with expansion along the hexagonal ab-plane with $a=a_0+3\%a_0$ and $c=c_0$, where $a_0,~c_0$ denote experimental lattice constants, does the job as shown in Fig.~2(c). The lattice expansion is seen to push the $s-$states well below the $p-$states, and to also open an insulating gap by lifting the four-fold degeneracy of $j=3/2$ states through the loss of cubic symmetry. Since the band inversion occurs only at one point in the Brillouin zone, non-cubic Li$_2$AgSb is a 3D strong topological insulator. We return to this point below to show more rigorously that the aforementioned band inversion correctly indicates that distorted non-cubic Li$_2$AgSb is topologically nontrivial.

\begin{figure}
\includegraphics[width=8.5cm]{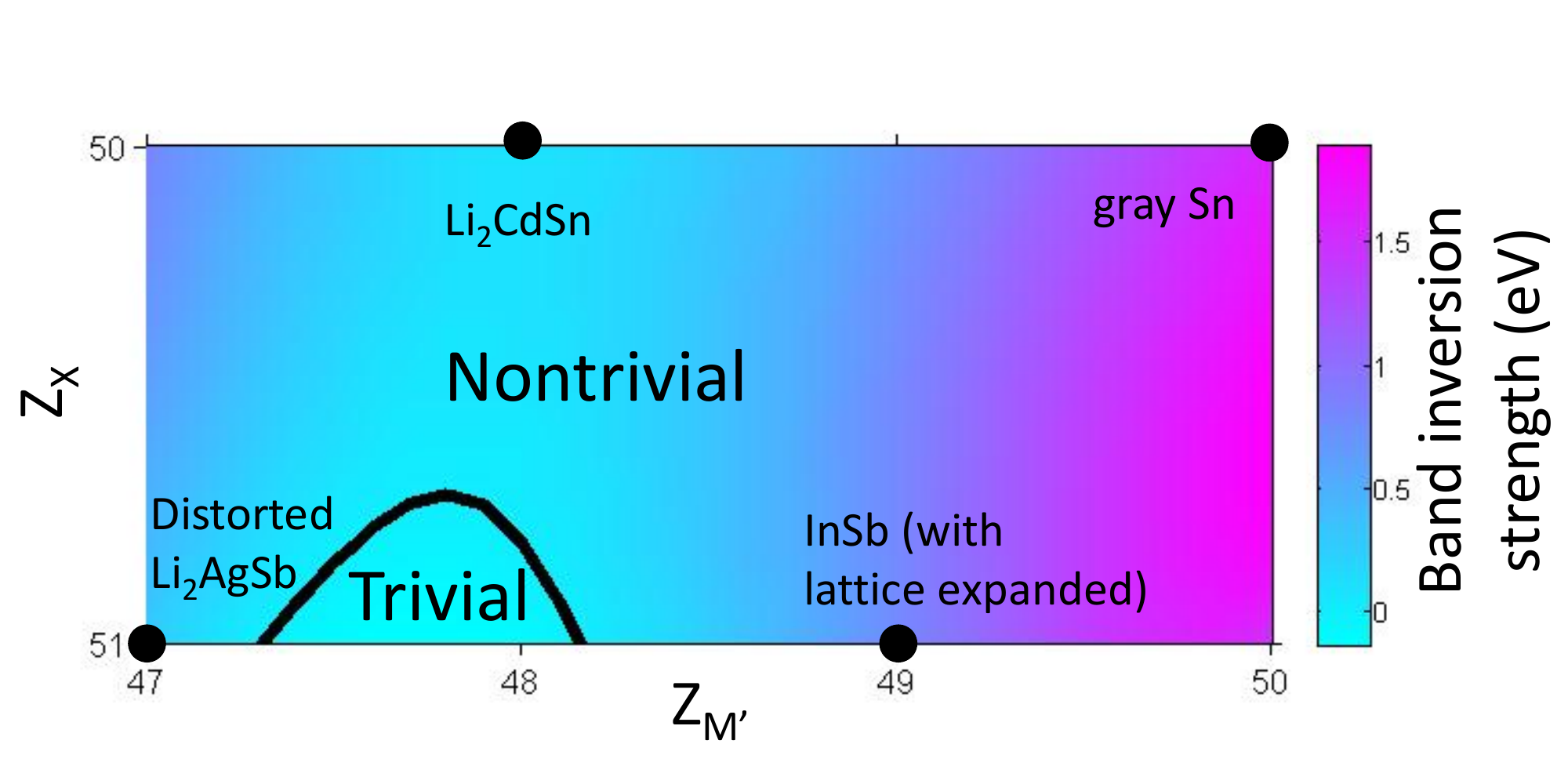}
\caption{(Color online) {Band inversion strength and topological phase diagram.}
Band inversion strength is plotted as a function of the atomic numbers of $M'$ and $X$ elements in the zinc-blende sublattice. The calculation is carried out at $a$ = 12.8 Bohr which is 3\% greater than the experimental lattice constant of Li$_2$AgSb. The black line, which defines the topological critical points
by the zero value of the band inversion strength, separates the trivial and non-trivial topological phases.}
\end{figure}

Next, we apply the adiabatic continuity principle to a larger range of atomic compositions of Li$_2M'X$ by systematically changing the nuclear charge $Z$ of the atoms. Let us assume that atoms at the Li, Ag, and Sb positions possess hypothetical nuclear charges $Z_{M}=3-0.5 x+0.5 y$, $Z_{M'}=47+x$, and $Z_X=51-y$, respectively, where $x$ and $y$ are adjustable parameters. This choice guarantees that the system remains neutral for all values of $x$ and $y$. Note that $x$ and $y$ need not be integers and that non-integral nuclear charges are easily accommodated in first-principles all-electron computations in order to change the Hamiltonian continuously, while maintaining charge self-consistency throughout the transformation process.  This mapping can be started with $x=0$ and $y=0$, which corresponds to Li$_2$AgSb, and end with $x=3$ and $y=1$, which corresponds to the artificial compound He$_2$SnSn. He$_2$SnSn possess inversion symmetry and the wavefunction parity analysis can be used to obtain the $Z_2$ topological invariant\cite{FuKane}. Since He is
chemically inert, He$_2$SnSn has similar band structure as Sn and is a topologically non-trivial semimetal with $Z_2=1$.

A topological phase diagram can now be constructed based on the band inversion strength as a function of atomic numbers, as shown in Fig.~3 in the region of $0\leq x\leq 3$ and $0\leq y\leq 1$ at the lattice constant $a$=12.8 Bohr, which is 3\% longer than the experimental value for Li$_2$AgSb. A borderline is drawn at the zero of the band inversion strength (thick black line). A phase transition occurs when one crosses this borderline in that the two regions separated by the borderline are not topologically equivalent.  Li$_2$AgSb with lattice expansion and gray tin are seen to lie on the same side of the thick black line, and are thus adiabatically connected, implying that they are topologically equivalent. Since $Z_2=1$ for gray tin (actually, He$_2$SnSn with lattice expansion or distortion), Li$_2$AgSb with lattice expansion [Fig.~2(b)] or distortion [Fig.~2(c)] is also topologically nontrivial with $Z_2=1$. The borderline itself marks topologically critical phases. It would move toward positive $y$ direction if the lattice constants decreased. Li$_2$AgSb with experimental lattice constants [Fig.~2(a)] lies very close to this critical line. Note that the variation of $Z$s can also be considered as a doping effect in the spirit of the virtual crystal approximations.\cite{new1,new2,new3} Therefore, our topological phase diagram in Fig.~3 could be used also to guide search of topologically interesting electronic structures through alloying.

\begin{figure}
\includegraphics[width=8.5cm]{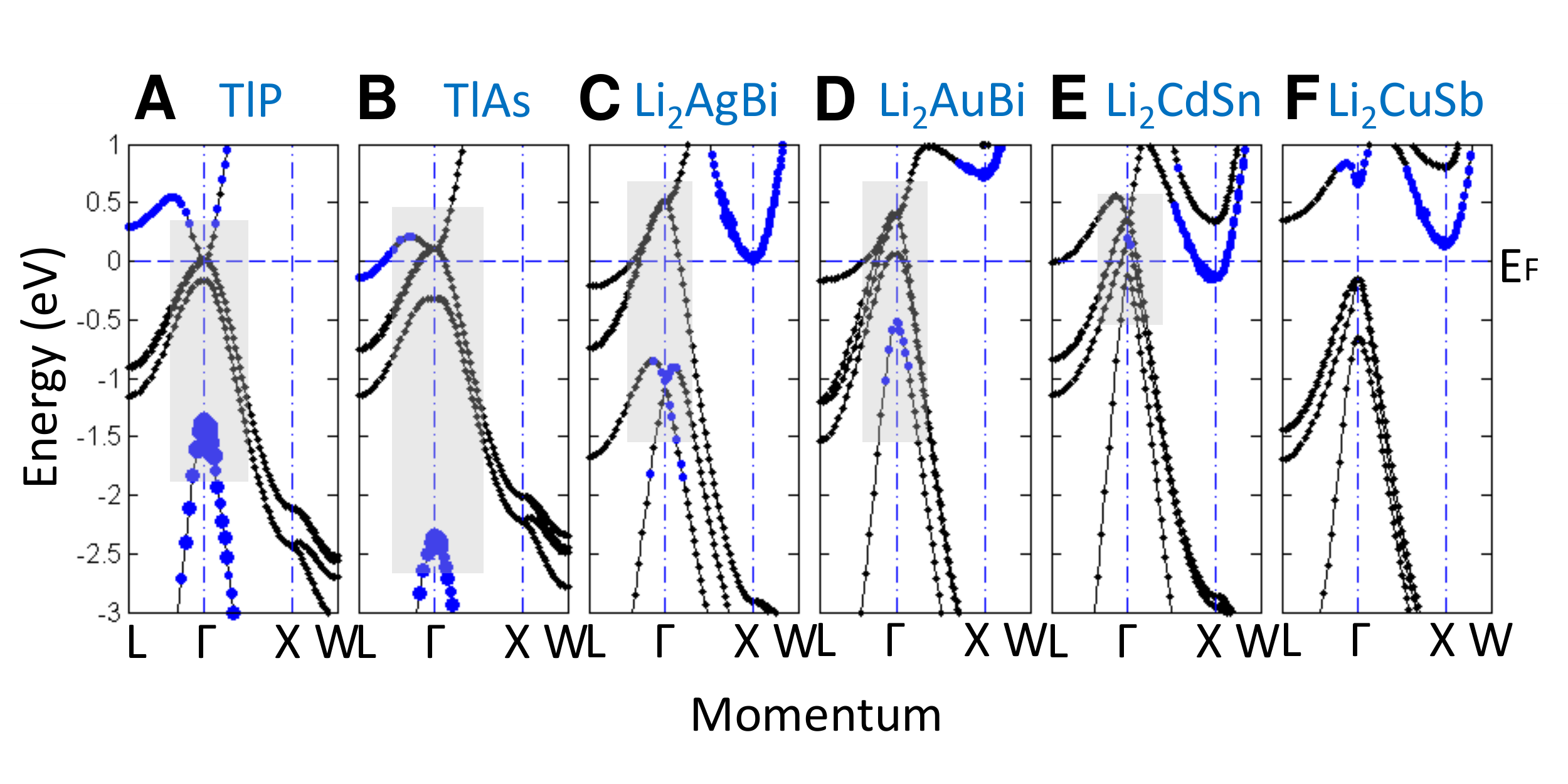}
\caption{(Color online) {Electronic structure of Li$_2M'X$ series and III-V binary compounds.}
Bulk electronic structures of TlP in (a), TlAs in (b), Li$_2$AgBi in (c), Li$_2$AuBi in (d), Li$_2$CdSn in (e), and Li$_2$CuSb in (f). The size of blue data points is proportional to the probability of s-orbital occupation on the anion site. Grey shaded areas in panels (a)-(e) highlight band inversion corresponding to the band-structure in Fig. 1(e).}
\end{figure}

To further establish the usefulness of the adiabatic continuation approach in extending the search for topologically non-trivial materials, we carefully examine the phase diagram of Fig. 3 in which we can identify two new candidates for topologically nontrivial materials, namely, InSb and Li$_2$CdSn. InSb is a topologically trivial semiconductor in its native phase with the experimental gap of 0.26 eV \cite{InSbgap}.  Here we predict that in InSb with adequate lattice expansion, the gap can be closed and a band inversion can occur. It is well known that the band gap in the III-V zinc-blende structure decreases
for heavier constituent atoms and the lattice constants are larger. Earlier first-principles calculations have predicted that among heavy-atom III-V compounds, TlP and TlAs are stable in the zinc-blende structures\cite{TlAs}. We take the total-energy optimized lattice constants\cite{TlAslattice} and obtain the band structures for TlP and TlAs in Fig.~4. Both compounds possess the band inversion feature at the $\Gamma$ point.  Therefore, thallium-based III-V compounds are predicted to be topologically nontrivial.  In addition, the II-VI compound CdTe lies at $x=1$ and $y=-1$ in the phase diagram of Fig. 3. While CdTe is topologically trivial, the related compound HgTe is a known topologically nontrivial semimetal\cite{HgTe}. Li$_2$CdSn is also reported to be a cubic crystal\cite{Li2CdSn}. If we assume Li$_2$CdSn has the same crystal structure as Li$_2$AgSb, the optimized lattice constant is obtained as $a_0=12.85$ Bohr. Our band structures of Li$_2$CdSn in Fig.~4 show that a band inversion occurs at the $\Gamma$ point and the conduction bands are below the $E_F$ around the $X$-point, allowing us to predict that Li$_2$CdSn is a topologically nontrivial metal.

Along the preceding lines, we can theoretically predict still other topological insulators. If we replace Li$_2$AgSb with heavier atoms, the band inversion can be achieved at the $\Gamma-$point (Fig.~4). We predict thus that Li$_2$AgBi and Li$_2$AuBi are non-trivial topological metals in their pristine phases whereas Li$_2$CuSb is a trivial band insulator. Another novel way of breaking the cubic symmetry of the zinc-blende structure to attain the topological insulating state is to double the unit cell, which leads to a tetragonal crystal structure. The chalcopyrite semiconductors of I-III-VI$_2$ and II-IV-V$_2$ group compositions which are described by this tetragonal structure, are shown to be non-trivial topological insulators in Ref.~\onlinecite{Fengchalcopyrite}, which is consistent with our predictions. Furthermore, we have found that quaternary chalcogenides and ternary famatinites, which are tetragonal with zinc-blende sublattice, are candidates for 3D topological insulators.\cite{Ray2011} We note that the adiabatic continuation method used in the present calculations also predicts that half-Heusler compounds are topologically non-trivial.\cite{heuslerhasan}

Concerning the practical realization of topologically nontrivial phases in Li$_2$AgSb class of compounds, as shown in Fig.~2(c), the nontrivial topological insulating phase is predicted to exist when a uniaxial tensile strain is applied. This could be achieved by growing the materials on suitable semiconducting or insulating substrates. Notably, all the materials discussed in the present study display electronic structures similar to those in HgTe/CdTe,  which were used to achieve the two dimensional topological insulating phase or the quantum spin Hall (QSH) phase in a quantum well system.\cite{HgTe} Therefore, our work also expands the list of candidate materials for realizing QSH phases.

In conclusion, we have predicted new classes of topological insulator materials in III-V semiconductors and Li-based compounds Li$_2M'X$. 
Small-gap semiconductors such as InSb could be turned into nontrivial phases with a lattice expansion.
With the example of Li$_2$AgSb, we show how the nontrivial insulating phase could be obtained by a uniaxial strain. 
The exciting opportunities present within the ternary Li$_2M'X$ series (with a number of variants) offer a new platform for realizing multi-functional topological devices for spintronics and fault-tolerant quantum computing applications. Undergirding our study is the demonstration that a first-principles approach based on arguments of adiabatic continuation provides a powerful tool for materials discovery in search for new topologically interesting materials encompassing centrosymmetric as well as non-centrosymmetric crystal structures. The topological phase diagrams of the sort we have obtained for the Li-based compounds in this work, which encode global materials characteristics extending from single element to binary, ternary and possibly quaternary compositions would provide systematic opportunities for materials discovery and tunability between trivial and non-trivial topological phases, and serve to motivate further research for both fundamental and applied purposes.

{\bf Acknowledgements.}
The work at Northeastern and Princeton is supported by
the Division of Materials Science and Engineering, Basic
Energy Sciences, U.S. Department of Energy Grants DE-FG02-07ER46352,
DE-FG-02-05ER46200 and AC02-05CH11231, and benefited from theory support at the Advanced Light Source, Berkeley, and the 
allocation of supercomputer time at NERSC and Northeastern University's
Advanced Scientific Computation Center (ASCC).
M.Z.H is supported by NSF-DMR-1006492 and DARPA-N66001-11-1-4110.


\begin{thebibliography}{99}
\bibitem{reviewHasan}
M. Z. Hasan, and C. L. Kane, 
{Rev. Mod. Phys.} {\bf 82}, 3045-3067 (2010).
%
\bibitem{reviewZhang}
X.-L. Qi, and S.-C. Zhang,  
{Rev. Mod. Phys.} {\bf 83}, 1057 (2011).
%
\bibitem{reviewMoore}
M. Z. Hasan, and J. E. Moore, 
{Annual Review of Condensed Matter Physics} {\bf 2}, 5578 (2011).
%
\bibitem{wilchek}
F. Wilczek, Journal Club. {Nature} {\bf 458}, 129 (2009).
%
\bibitem{ZhangDyon}
X.-L. Qi, T. L. Hughes, and S.-C. Zhang, 
{Phys. Rev. B} \textbf{78}, 195424 (2008).
%
\bibitem{essinaxion}
A. M. Essin, J. E. Moore, and D. Vanderbilt, 
{Phys. Rev. Lett.} {\bf 102}, 146805 (2009)
%
\bibitem{exciton}
B. Seradjeh, J. E. Moore, and M. Franz,
{Phys. Rev. Lett.} {\bf 103}, 066402 (2009).
%
\bibitem{Zhangmonopole}
X.-L. Qi, R. Li, J. Zhang, and S.-C. Zhang, 
{Science} {\bf 323}, 1184 (2009).
%
\bibitem{Majorana}
L. Fu, and C. L. Kane, 
{Phys. Rev. Lett.} \textbf{100}, 096407 (2008).
%
\bibitem{WrayCuBiSe}
L. A. Wray {\it et al.},
{Nat. Phys.} \textbf{6}:855-859.
%
\bibitem{MatthewNatPhys}
Y. Xia  \textit{et al.},
{Nat. Phys.} \textbf{5}, 398-402 (2009).
%
\bibitem{BiTeSbTe}
D. Hsieh \textit{et al.}, 
{Phys. Rev. Lett.} \textbf{103}, 146401 (2009).
%
%
\bibitem{tetradymite}
H. Lin {\it et al.},
{New Journal of Physics} {\bf 13}, 095005 (2011).
%
\bibitem{heuslerhasan}
H. Lin {\it et al.} 
{Nature Materials}  {\bf 9}, 546 (2010)
%
\bibitem{heuslerZhang}
S. Chadov {\it et al.} Nature Materials \textbf{9}, 541 (2010).

\bibitem{TlBiTe2}
H. Lin {\it et al.}, 
{Phys. Rev. Lett.} \textbf{105}, 036404 (2010).
%

\bibitem{TlBiTe2Zhang}
B. Yan {\it et al.}, 
EPL (Europhysics Letters) \textbf{90}, 37002 (2010).

\bibitem{SuYangThallium}
S.-Y. Xu {\it et al.}, 
{Science} {\bf 332}, 560-564 (2011).
%
\bibitem{SuYangGBT}
S.-Y. Xu {\it et al.},
arXiv:1007.5111.
%
\bibitem{Ray2011}
Y. J. Wang {\it et al.}
{New Journal of Physics} {\bf 13}, 085017 (2011).
%
\bibitem{Tkondo}
M. Dzero, K. Sun, V. Galitski, and P. Coleman, 
{Phys. Rev. Lett.} {\bf 104}, 106408 (2010).
%
\bibitem{TAFM}
R. S. K. Mong, A. M. Essin, and J. E. Moore,  
{Phys. Rev. B} {\bf 81}, 245209 (2010).
\bibitem{TMott}
S. Raghu, X.-L. Qi, C. Honerkamp, and S.-C. Zhang, 
{Phys. Rev. Lett.} {\bf 100}, 156401 (2008).
%
\bibitem{THO}
T. Das, Sci. Rep. 2, 596 (2012).
%
\bibitem{topocrystal}
L. Fu, 
{Phys. Rev. Lett.} {\bf 106}, 106802 (2011).
%
\bibitem{FuKane}
L. Fu, and C. L. Kane,
{Phys. Rev. B} \textbf{76}, 045302 (2007).
%
%
\bibitem{wien2k}
P. Blaha {\it et al.}
\textit{WIEN2k, An augmented plane wave plus local orbitals program for calculating crystal properties.} (Karlheinz Schwarz, Techn. University Wien, Austria) (2001).
%
\bibitem{PBE96}
J. P. Perdew, K. Burke, M. Ernzerhof,
{Phys. Rev. Lett.} \textbf{77}, 3865-3868 (1996).
%
\bibitem{new1} H. Lin, S. Sahrakorpi, R.S. Markiewicz, and A. Bansil, Phys. Rev.
Lett. \textbf{96}, 097001 (2006); A. Bansil, Phys. Rev. B \textbf{20}, 4025 (1979); A. Bansil, Phys. Rev. B \textbf{ 20}, 4035 (1979). 
%
\bibitem {new2} It will be interesting to examine effects of alloying and doping using first-principles approaches beyond the virtual crystal or rigid band type approximations, see Ref.~\onlinecite{new3}. 
%
\bibitem{new3} A. Bansil, R. S. Rao, P. E. Mijnarends and L. Schwartz, Phys. Rev. B \textbf{23}, 3608 (1981); P. E. Mijnarends and A. Bansil, Phys. Rev. B \textbf{13}, 2381 (1976); L. Huisman, D. Nicholson, L. Schwartz and A. Bansil, Phys. Rev. B \textbf{24}, 1824 (1981); L. Schwartz and A. Bansil, Phys. Rev. B \textbf{10}, 3261 (1974).
%
\bibitem{InSbgap}
Sher van Schilfgaarde, and A.-B.Chen, 
{Appl. Phys. Lett.} \textbf{62}, 1857-1859 (1993).
%
\bibitem{TlAs}
Y. O. Ciftci, K. Colakoglu, E. A. Deligoz, 
{Cent. Eur. J. Phys.} \textbf{6}, 802-807 (2008).
%
\bibitem{TlAslattice}
N. Saidi-Houat, A. Zaoui, and M. Ferhat, 
{J. Phys.: Condens. Mat.} {\bf 19}, 106221  (2007).
%
\bibitem{HgTe}
B. A. Bernevig, T. L. Hughes, and S.-C. Zhang, 
{Science} {\bf 314}, 1757 (2006).
%
\bibitem{Li2CdSn}
H.-U. Schuster, {\it Naturwissenschaften} \textbf{53}, 361 (1966).
%
\bibitem{Fengchalcopyrite}
W. Feng, D. Xiao, J. Ding, and Y. Yao, {Phys. Rev. Lett.} {\bf 106}, 016402 (2011).
%
\end{thebibliography}
\end{document}